\documentclass[12pt]{report}
\usepackage[utf8]{inputenc}
\usepackage{blindtext}
\usepackage[letterpaper, total={8.5in, 11in}, margin=2in]{geometry}
\usepackage{ragged2e}
\usepackage{blindtext}
\usepackage{graphicx}
\usepackage{amsmath}
\usepackage{amssymb}
\usepackage{gensymb}
\usepackage{array}

\date{}

\begin{document}
\centering{\huge IRS for Multi-Access Edge Computing in 6G Networks\\
\vspace{24pt}
\large Mobasshir Mahbub, Raed M. Shubair}

\newpage

\RaggedRight{\textbf{\Large 1.\hspace{10pt} Introduction}}\\
\vspace{18pt}
\justifying{\noindent The last three decades have witnessed an exponential growth and tremendous developments in wireless technologies and techniques, and their associated applications.  These include indoor localization techniques and related aspects [1-17], terahertz communications and signal processing applications [18-36], and antenna design and propagation characteristics [32-55].\par
Mobile data will rise exponentially in the forthcoming sixth-generation (6G) [56], [57] infrastructures and the majority of it is generated by edge devices including smartphones, computers, and Internet of Things (IoT) devices [58]. Since the majority of such user equipment holds low computational capacity, MEC is a potential technology that may efficiently guarantee sufficient computing resources, minimize capital expenditure, and provide scalability [59]. To preserve battery power and computing resources, wireless gadgets in the MEC architecture transfer computing-intensive or delay-sensitive operations to adjacent edge computing servers located at the edge of cellular networks.\par
MEC encourages the deployment of cloud processing technologies at the edge of wireless networks to relieve resource-constrained UEs from excessive computational workloads and serve them with highly-efficient low-latency computational services [60].\par
Because of the potential to change the wireless propagation environment, IRS has already been anticipated as a revolutionary approach 6G connectivity system [61], [62]. IRS is comprised of a large number of reflecting (passive) components that may proactively reflect incident waves to boost signal strength or minimize co-channel interference by altering their phases and amplitudes. Unlike conventional relay transmissions, IRS not only functions in a full-duplex manner without introducing unwanted interference, but also it can significantly minimize energy consumption and hardware costs by utilizing passive reflecting materials. Fig. 1 illustrates the IRS-assisted communication scenario.
\vspace{6pt}

\begin{figure}[h]
    \centering
    \includegraphics[height=7.5cm, width=10.5cm]{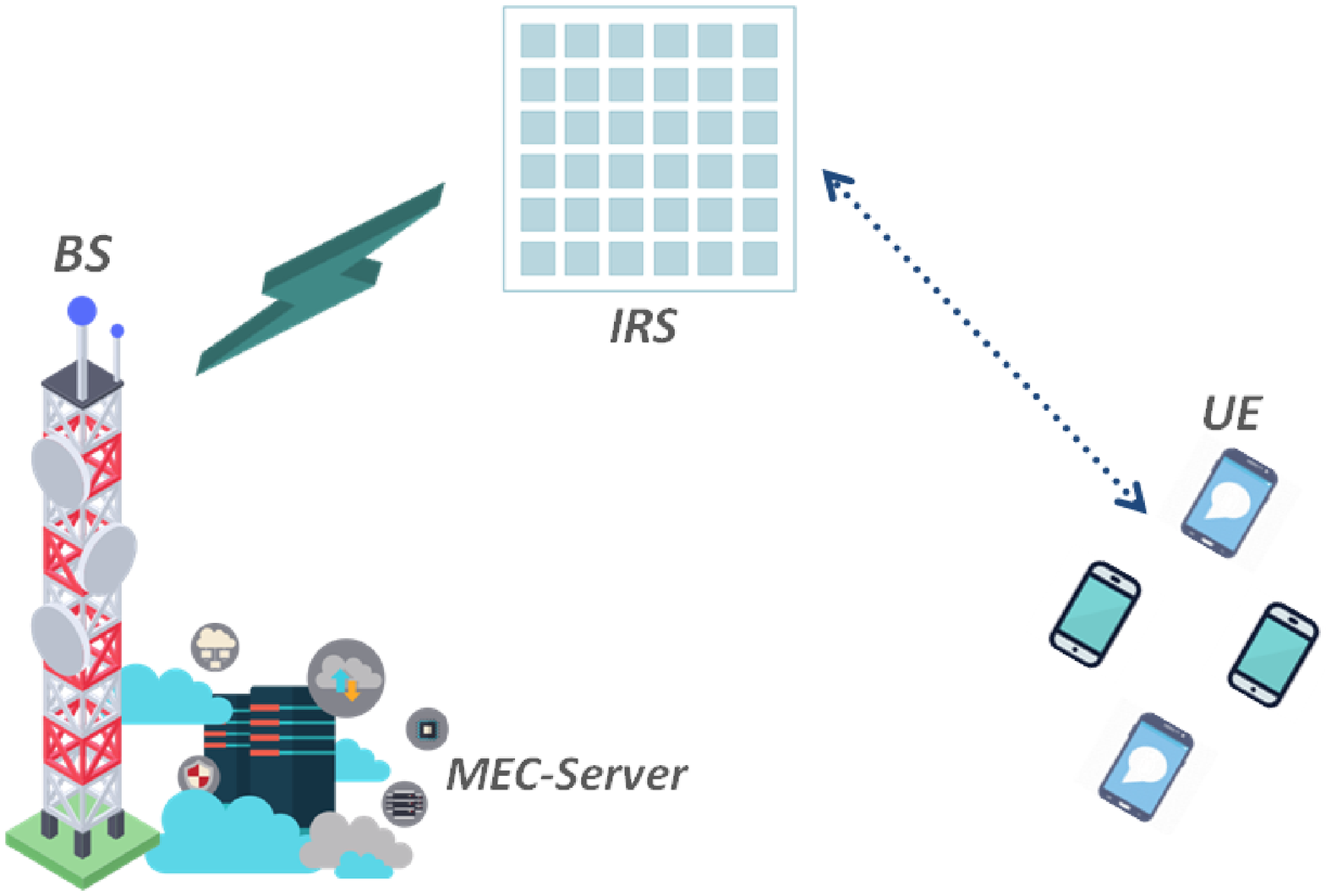}\\
    Fig. 1. IRS-empowered UAV-assisted wireless network
\end{figure}
IRS technology has already attracted significant consideration to improve the offloading efficiency of resource-constrained UE because of its benefits of simple deployment, low cost, smart-controllable passive beamforming, and signal improvement or interference reduction [63]. The configurations of IRSs may be modified to offer a more advantageous wireless transmission environment by altering the characteristics of reflecting components on the surface. Clearly, including IRSs into MEC technology [64] is a cost-efficient and environmentally benign method of offloading the computational tasks of UE.\par
The research aimed to analyze the performance of MEC considering in both IRS-assisted and without IRS communication scenarios.\par}
\vspace{18pt}

\RaggedRight{\textbf{\Large 2.\hspace{10pt} Relative Literature}}\\
\vspace{18pt}
\justifying{\noindent The section briefed several prior research works relative to IRS-empowered MEC systems.\par
Wang et al. [65] studied an IRS-empowered MEC adopting the non-orthogonal multiple access (NOMA) technique. The energy efficiency is jointly optimized considering the offloading power, receiving beamforming, local computing frequency, and IRS phase-shift. Zhou et al. [66] considered IRS-assisted MEC architecture in which IRS is employed to support computation task offloading from users (two users) to an access node linked with an edge computing cloud. Chu et al. [67] deployed and analyzed the computational performance of IRS-enhanced MEC architecture. The work formulated a problem targeting to optimize sum computational bits considering the offloading time allocation, CPU frequency, transmit power of each device, etc. Zhang et al. [68] investigated the network throughput optimization problem of IRS-aided multiple hop MEC system. The work jointly optimized resource allocation and phase-shifts of the IRS. Bai et al. [69] investigated the lucrative role of IRS in MEC architecture in which single-antenna user equipment may offload partial computational tasks or activities to the computing node (edge computing) via a multi-antenna access node with the assistance of an IRS. Latency-minimization problems are composed for both multi-device and single-device scenarios. Mao et al. [70] aimed at utilizing the IRS to escalate the efficiency of wireless energy transfer and task offloading. Specifically, the work investigated the maximization of total computation bits for IRS-assisted wireless powered MEC systems through the joint optimization of the transmission power, beamforming of IRS, and time slot assignment. Sun et al. [71] presented a new IRS-MEC framework that jointly optimized the local computing frequencies (CPU cycles) of the UE and the offloading schedules to reduce the consumption of energy of the UE.}\par
\vspace{18pt}

\RaggedRight{\textbf{\Large 3.\hspace{10pt} System Model}}\\
\vspace{12pt}

\justifying{\noindent The research considered a MEC-enabled wireless network scenario. $M = \{M_1,M_2,… M_n\}$ where i is the serving base station (BS) from M. A set $U = \{u_1,u_2,… u_n\}$ of the user equipment is served by the corresponding BS i. User equipment operating under the BS has to perform multi-varieties of computational tasks such as multiplayer gaming, facial recognition or verification, image processing or analysis, etc. The computational or computing tasks of users are indicated by $T_n   = \{d_n,c_n,t_n^{max}\}$ where $n \in i$. $d_n$ is indicating the computation task data packet size, $c_n$ denotes the required CPU cycles to complete each task, and $t_n^{max}$ indicates the tolerable latency (maximum) of each task.}\par
\vspace{12pt}
\RaggedRight{\textit{\large A.\hspace{10pt} Communication Model}}\\
\vspace{12pt}
\justifying{\noindent \textbf{Communication Model without IRS:} Contemplate $P_t^{UL}$ and $B_t^{UL}$ denote the power and bandwidth for uplink [72].\par
The uplink received power (by the BS) can be measured by (Eq. 1),
}
\vspace{6pt}
\begin{equation}
P_r^{UL}= \frac{P_t^{UL}h}{D^\alpha} 
\end{equation}
where $h$ is unit mean Rayleigh fading coefficient that evolves exponential distribution $h \sim exp(1)$. $D$ denotes the separation distance between transmitter and receiver. The path loss factor is denoted by $\alpha$.\par
The SNR for uplink can be determined by the following formula (Eq. 2),
\vspace{6pt}
\begin{equation}
S_r^{UL}= \frac{P_r^{UL}}{N}
\end{equation}
\vspace{6pt}
where $N$ indicates the interference power.\par
Therefore, the uplink throughput can be measured by (Eq. 3),
\vspace{6pt}
\begin{equation}
R^{UL}= B_t^{UL}log_2(1+\frac{P_r^{UL}}{N})
\end{equation}
\vspace{6pt}

\justifying{\noindent \textbf{IRS-Assisted Communication Model:} In the IRS-assisted communication, the received power (uplink) can be derived by the following equation (Eq. 4) [73],}\par
\begin{equation}
P_r^{UL(IRS)} = \frac{P_t^{UL(IRS)}G_t G_r G M^2 N^2 d_x d_y \lambda^2 cos(\theta_t) cos(\theta_r) A^2}{64\pi^3(d_1 d_2)^2}
\end{equation}
\vspace{6pt}
where $P_t^{UL(IRS)}$ is the transmit power of the user equipment in an IRS-aided communication scenario.\par
\noindent $d_1=  \sqrt{(x^{UE}-x_i^{IRS} )^2+(y^{UE}-y_i^{IRS} )^2+(z^{UE}-z_i^{IRS} )^2}$ indicates the distance between the UE located at $(x^{UE},y^{UE},z^{UE})$ coordinates and IRS located at $(x_i^{IRS},y_i^{IRS},z_i^{IRS})$ coordinates. $d_2=  \sqrt{(x_i^{IRS}-x_n^{BS} )^2+(y_i^{IRS}-y_n^{BS} )^2+(z_i^{IRS}-z_n^{BS} )^2}$
is the distance between the IRS and the BS. The transmitter and receiver gains are $G_t$ and $G_r$. $G=\frac{4\pi d_x d_y}{\lambda^2}$  denotes the scattering gain. M and N indicate the numbers of transmit-receive elements in IRS. The length and width of scattering elements of the IRS are denoted by $d_x$ and $d_y$. $\lambda$ is the wavelength. Transmit and receive angles are $\theta_t$ and $\theta_r$. $A$ is the amplitude respective to the reflection coefficient of IRS.\par
The SNR is calculated by the formula below (Eq. 5),
\vspace{6pt}
\begin{equation}
S_r^{UL(IRS)}= \frac{P_r^{UL(IRS)}}{N}
\end{equation}
\par
\vspace{6pt}
The uplink throughput is measured by (Eq. 6),
\vspace{6pt}
\begin{equation}
R^{UL(IRS)}= B_t^{UL(IRS)}log_2(1+\frac{P_r^{UL(IRS)}}{N})
\end{equation}
where $B_t^{UL(IRS)}$ is the uplink bandwidth of the user equipment in the IRS-assisted scenario.\par
\vspace{12pt}
\RaggedRight{\textit{\large B.\hspace{10pt} Computation Model}}\\
\vspace{12pt}
\justifying {\noindent \textbf{Local Computation:} Contemplating $F^L$ is the maximum operable frequency of the CPU of user equipment. The latency of task processing (time for processing a task by user equipment as per the computation capacity) in the local computing model can be derived by (Eq. 7) [74],
\vspace{6pt}
\begin{equation}
t_n^{L}= \frac{d_n c_n}{F^L - \sum_1^n f_n^L}
\end{equation}
where $f_n^L$ denotes the occupied CPU frequency by the previously ongoing task/tasks.}\par
\vspace{12pt}
\justifying {\noindent \textbf{Computation at the BS (without IRS):} Based on the task processing computation capacity, time (latency) constraint, and power consumption or limitation of battery level the user equipment can have a decision to offload computational task/tasks to the MEC server. In this circumstance, the task completion time $T_n$ can be partitioned into two portions; i.e. one portion denotes the task offloading or transmission time or latency and another portion stand for task processing time at the MEC server. The work disregarded the time required for the reception or the computation results from the MEC to the UE since the derived result is usually much tinier in size than the input data size for most of the cases (e.g., fingerprint recognition). $F^{BS}$ is the maximum task processing capacity (CPU frequency) of the MEC. $\sum_1^n f_n^{BS}$  denotes the occupied capacity of the CPU of the MEC server. Therefore, the total time for the task completion can be measured by (Eq. 8),
\vspace{6pt}
\begin{equation}
t_n^{BS}= \frac{d_n}{R^{UL}} + \frac{d_n c_n}{F^{BS} - \sum_1^n f_n^{BS}}
\end{equation}}
\vspace{6pt}
\justifying {\noindent \textbf{IRS-Assisted Computation:} In the circumstance of IRS-assisted communication, the overall task completion time can be derived by (Eq. 9),
\vspace{6pt}
\begin{equation}
t_n^{BS(IRS)}= \frac{d_n}{R^{UL(IRS)}} + \frac{d_n c_n}{F^{BS} - \sum_1^n f_n^{BS}}
\end{equation}}
\vspace{18pt}

\RaggedRight{\textbf{\Large 4.\hspace{10pt} Results and Discussions}}\\
\vspace{12pt}
\justifying{\noindent The section of the paper includes the measurement results and corresponding discussions on the derived results. Table I includes the measurement parameters and values.\par
\vspace{12pt}
\begin{center}
\begin{tabular}{| m{5cm} | m{5cm}|}
\hline
\textbf {Simulation Parameters} & \textbf{Values}\\
\hline
Cell area & 200x200 m\\
\hline
UE transmission power &	5W (w/o IRS); 2W (IRS-aided)\\
\hline
Bandwidth & 1 - 10 MHz\\
\hline
Transmitter gain & 20 dB\\
\hline
Receiver gain & 20 dB\\
\hline
No. of transmit and receive elements of IRS & 100\\
\hline
Transmit and receive size & 0.0038 mm\\
\hline
Carrier frequency & 120 GHz\\
\hline
Transmit and receive angle & 45\degree\\
\hline
BS position & (0, 0, 8) m\\
\hline
IRS position & (100, 100, 8) m\\
\hline
UE-IRS and IRS-BS distance & 100 m\\
\hline
Path loss factor & 5.5\\
\hline
IRS reflection coefficient & 0.9\\
\hline
Data sizes & 5000 – 20000 B\\
\hline
CPU cycles per bit & 1000\\
\hline
UE computation capacity & 2-4 GHz\\
\hline
BS computation capacity & 80 GHz (max. 8 GHz per user)\\
\hline
\end{tabular}
\end{center}
The research considered 30 ms or 0.030 s as the task completion threshold for the considered data sizes.\par
Fig. 2 shows the numerical results of task completion time in terms of data size for different processing capacities (CPU frequency) of different user equipment.}\par
\vspace{6pt}

\centering
\includegraphics[height=9.0cm, width=11.5cm]{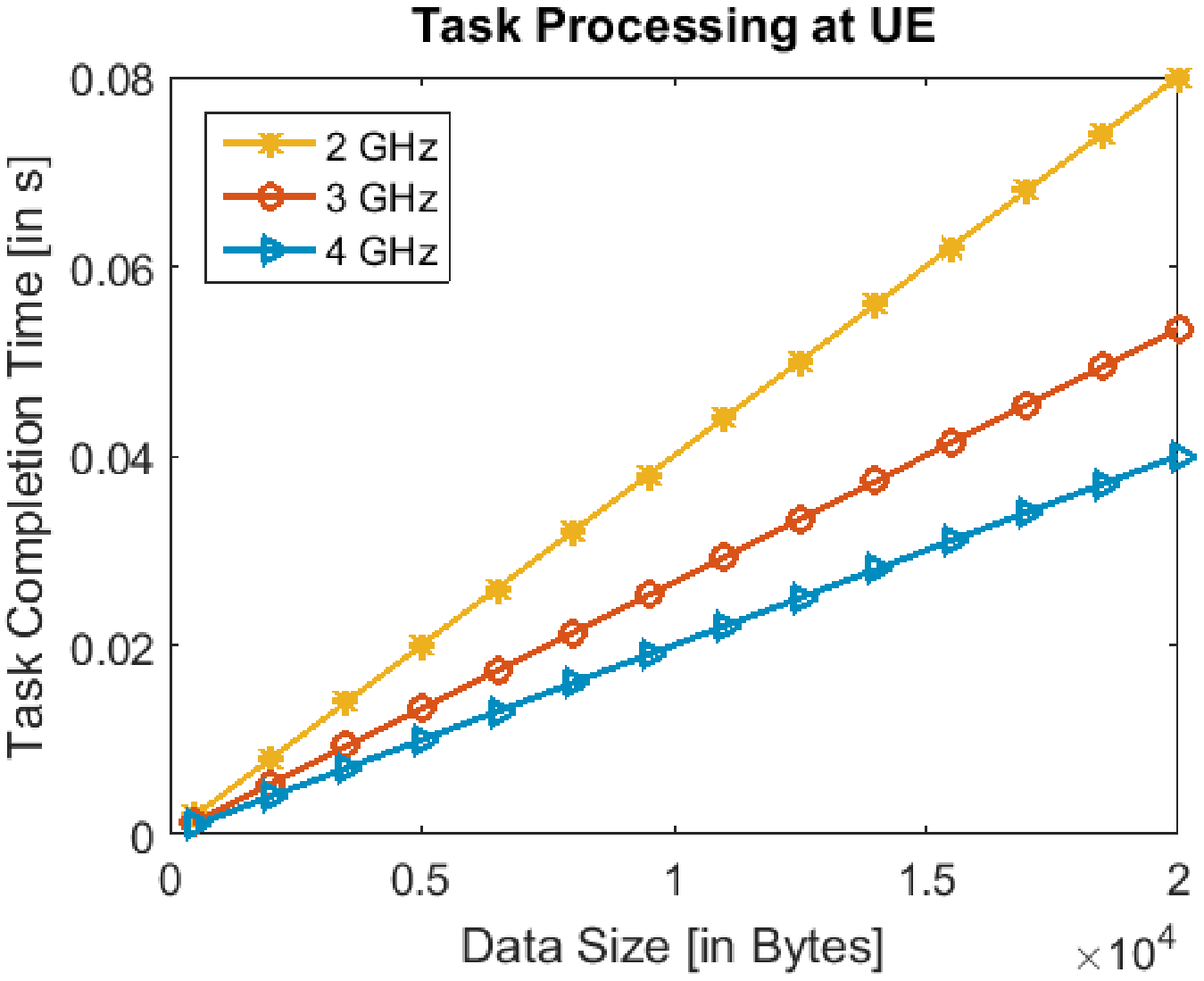}\par
\vspace{12pt}
Fig. 2. Task completion time vs. data size\par
\vspace{6pt}

\justifying{From Fig. 2 it is comprehensible that, UE with 2, 3, and 4 GHz computation or processing capacity can process up to 7500, 12000, and 16000 bytes of data size respectively within 30 ms task completion time (latency) threshold.}\par
\justifying{Fig. 3 illustrates the task completion time in terms of the allocated transmission bandwidth of the user equipment for both IRS-assisted and without IRS communication scenarios for 6000 bytes of data size.}
\vspace{6pt}

\centering
\includegraphics[height=9.0cm, width=11.5cm]{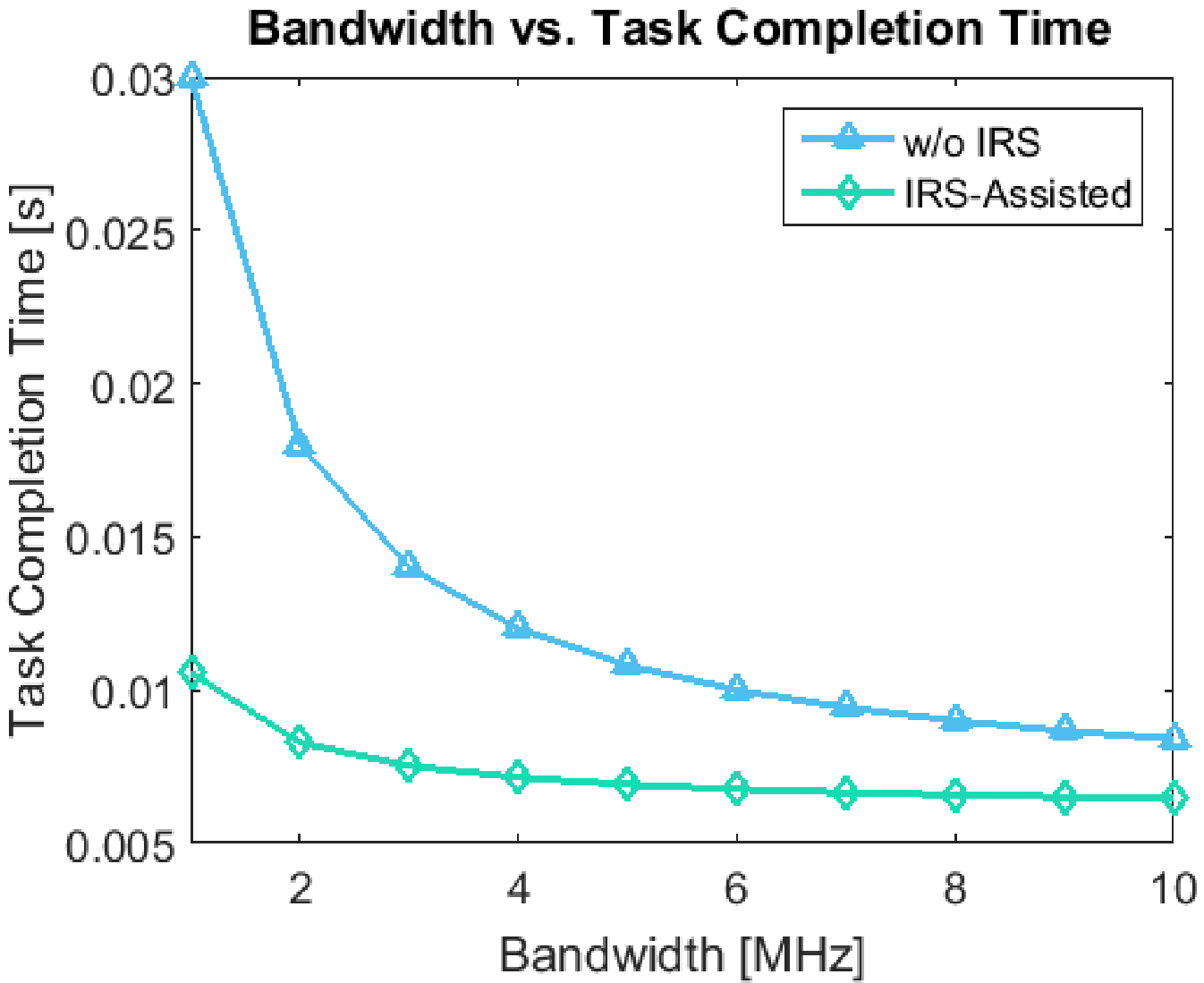}\par
\vspace{6pt}

Fig. 3. Task completion time vs. bandwidth (6000 B)
\vspace{6pt}

\justifying{Fig. 4 illustrates the task completion time relative to the bandwidth of the user equipment for both IRS-assisted and without IRS communications for 17000 bytes of data size.}
\vspace{6pt}

\centering
\includegraphics[height=9.0cm, width=11.5cm]{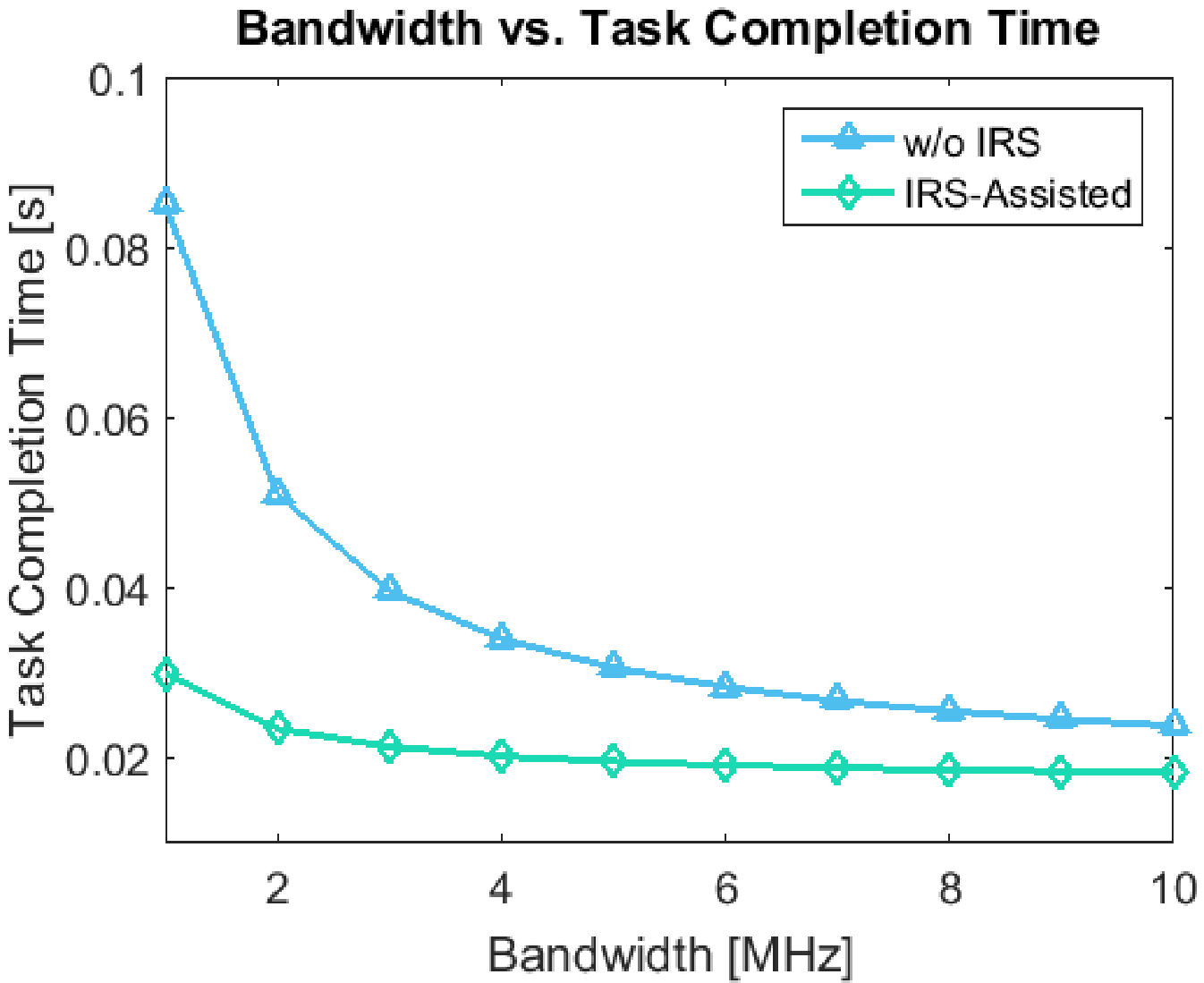}\par
\vspace{6pt}
Fig. 4. Task completion time vs. bandwidth (17000 B)
\vspace{6pt}

\justifying{From Fig. 3 it is perceptible that, in the case of non-IRS communication up to 6000 bytes of data can be offloaded and processed within the threshold task completion time in the case of minimal bandwidth (1 MHz). With such minimal bandwidth, up to 17000 bytes of data can be offloaded and processed within the threshold task completion time when IRS is deployed (Fig. 4). Almost 7 MHz bandwidth is required to offload and process 17000 bytes of data in the case of a non-IRS communication scenario.}\par
\justifying{Fig. 5 visualizes the task completion time relative to the bandwidth of the user equipment for both IRS-assisted and without IRS communications for 20000 bytes of data size.}
\vspace{6pt}

\centering
\includegraphics[height=9.0cm, width=11.5cm]{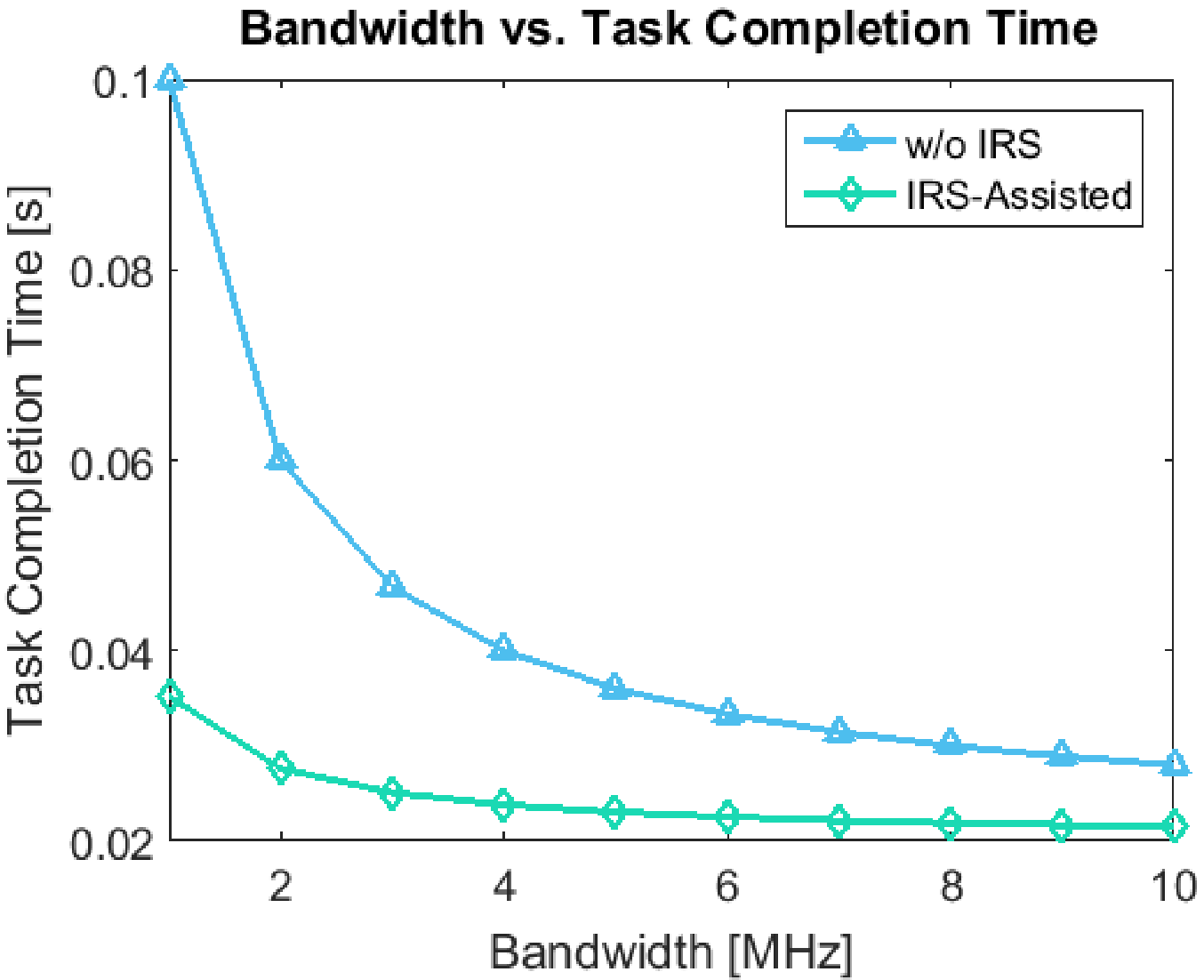}\par
\vspace{6pt}
Fig. 5. Task completion time vs. bandwidth (20000 B)
\vspace{6pt}

\justifying{With the observation of Fig. 5, it is sensible that, above 2 MHz bandwidth is required to offload 20000 bytes of data to be processed within the threshold task completion time. In the case of non-IRS communications, 8 MHz bandwidth is required to process 20000 bytes of data.}\par
\justifying{Fig. 6 represents the uplink throughput relative to the transmission bandwidth and transmitter-receiver separation in the context of without IRS micro cellular communication scenario.}
\vspace{6pt}

\centering
\includegraphics[height=9.0cm, width=11.5cm]{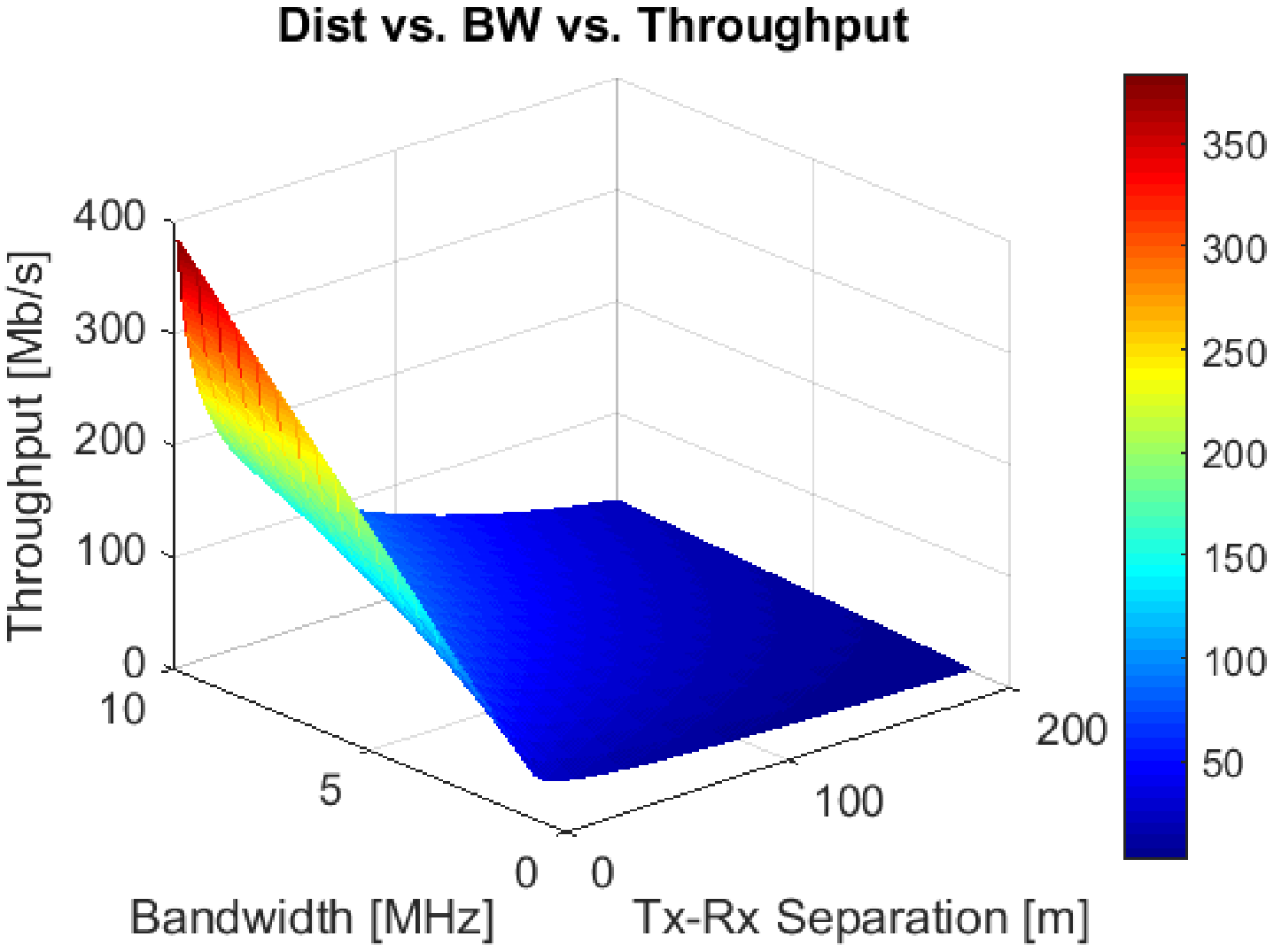}\par
\vspace{6pt}
Fig. 6. Uplink throughput in terms of bandwidth and transmitter-receiver separation (without IRS)
\vspace{6pt}

\justifying{Fig. 7 shows the task completion time relative to the transmission bandwidth and data size in the case of without IRS communication scenario.}
\vspace{6pt}

\centering
\includegraphics[height=9.0cm, width=11.5cm]{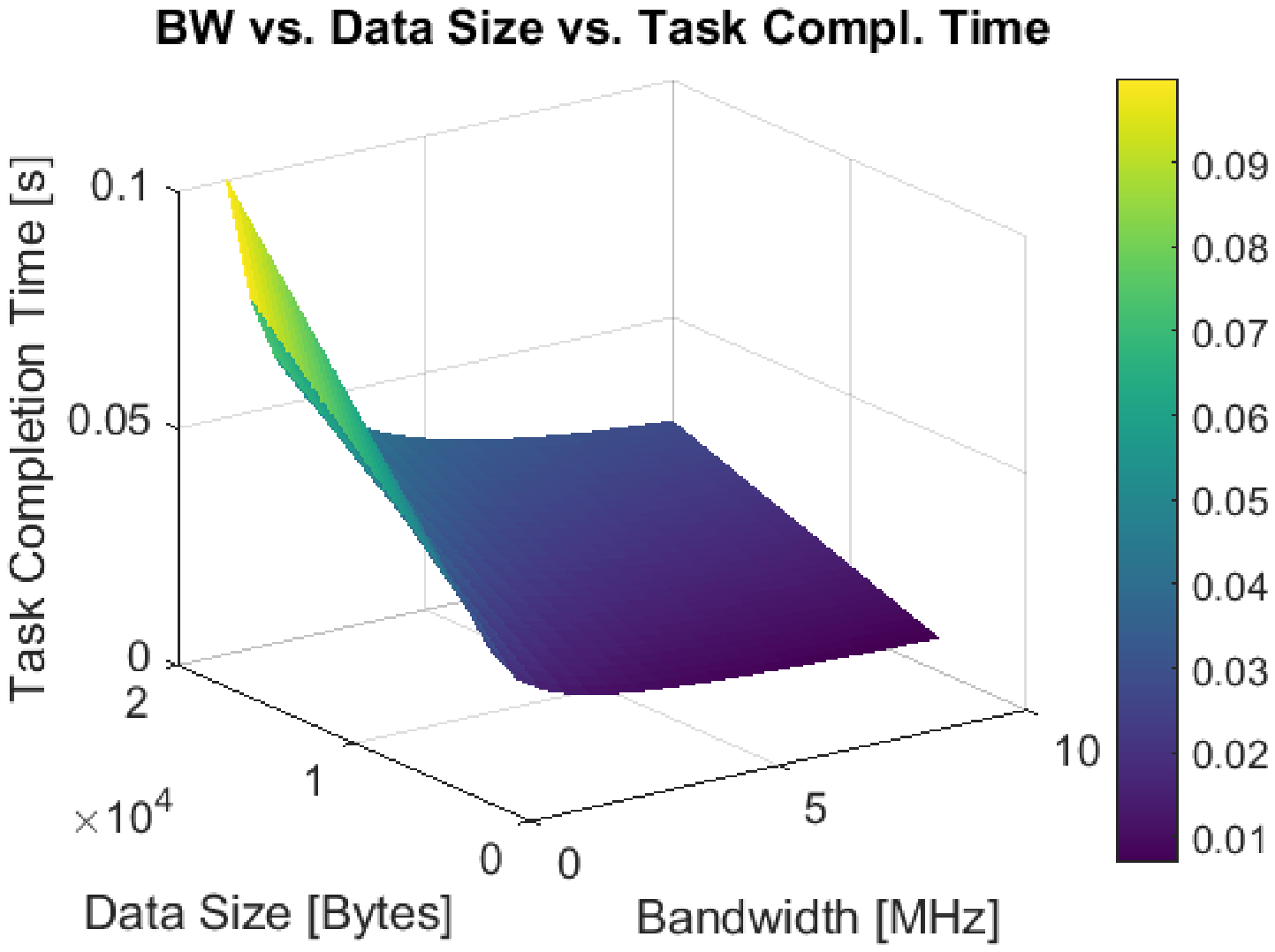}\par
\vspace{6pt}
Fig. 7. Task completion time in terms of bandwidth and data size (without IRS)
\vspace{6pt}

\justifying{Fig. 8 visualizes the uplink throughput relative to the bandwidth and transmitter-receiver separation distance in the context of an IRS-assisted micro cellular communication scenario.}
\vspace{6pt}

\centering
\includegraphics[height=9.0cm, width=11.5cm]{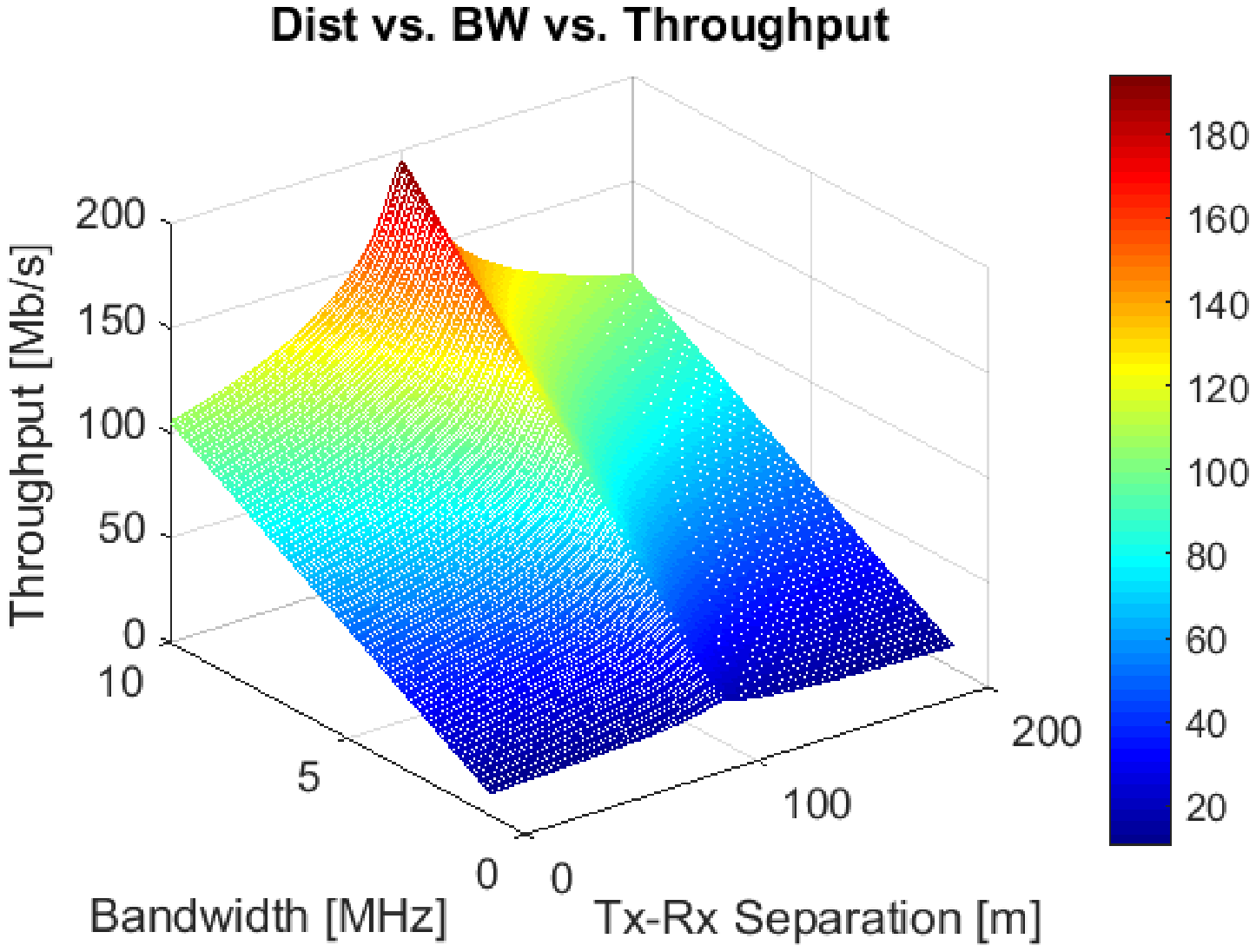}\par
\vspace{6pt}
Fig. 8. Uplink throughput in terms of bandwidth and transmitter-receiver separation (IRS-assisted)
\vspace{6pt}

\justifying{Fig. 9 illustrates the task completion time relative to the bandwidth and data size in the case of an IRS-assisted communication scenario.}
\vspace{6pt}

\centering
\includegraphics[height=9.0cm, width=11.5cm]{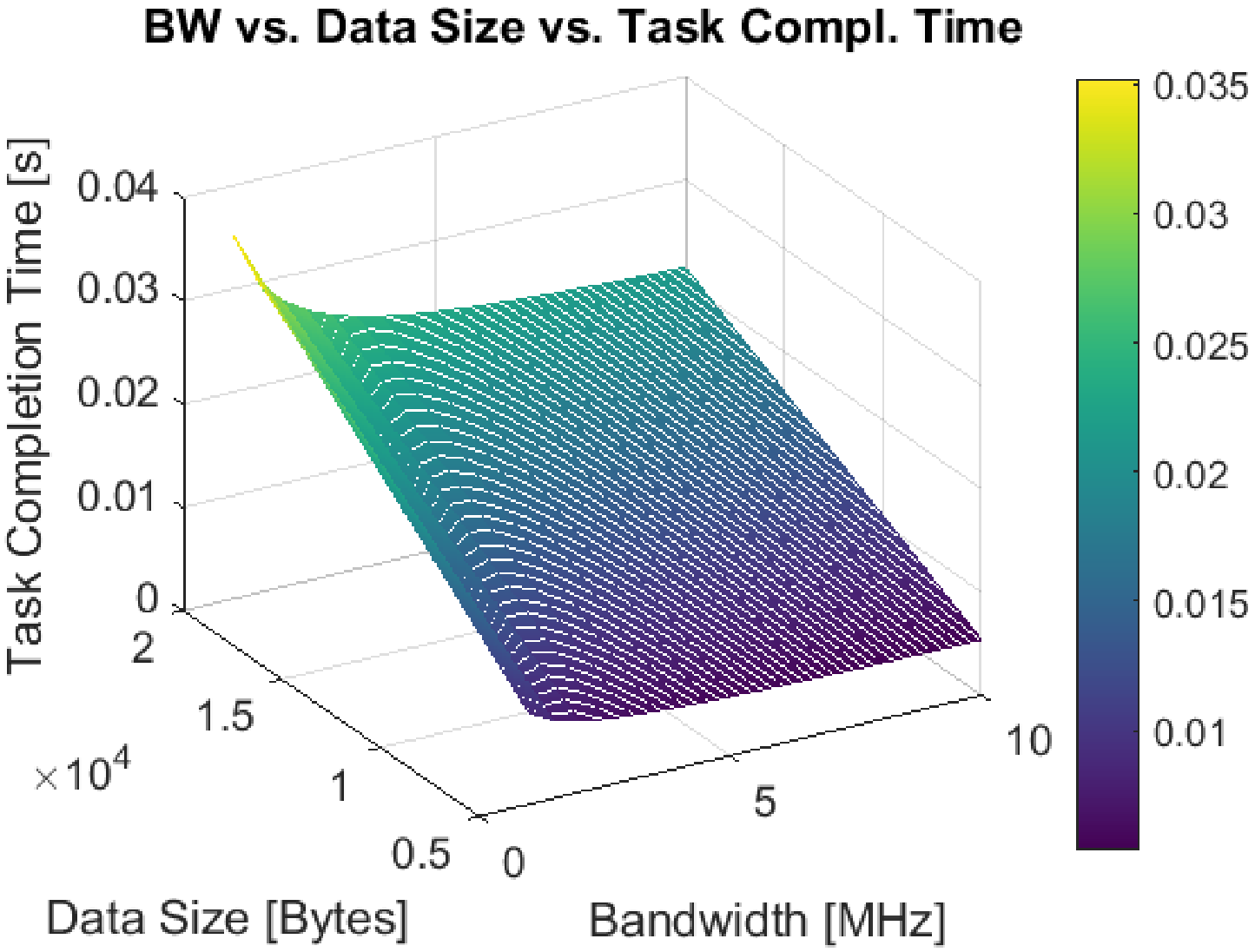}\par
\vspace{6pt}
Fig. 9. Task completion time in terms of bandwidth and data size (IRS-assisted)
\vspace{6pt}

\justifying{Through the observation and comparison between Figs. 6 and 8 it is comprehensible that, UE located at 200 m away from the BS for 1, 5, and 10 MHz bandwidths can obtain 2.001, 10.01, and 20.01 Mb/s uplink throughputs respectively in the case of non-IRS communication. On the other hand, in the case of IRS-assisted communication for the mentioned BS-UE separation distance UE can obtain 10.53, 52.63, and 105.3 Mb/s uplink throughputs for the respective mentioned bandwidths. It is evident that the deployment of IRS offers approximately 5 times increased uplink throughput.\par
Inspecting Figs. 7 and 9 it becomes sensible that, in the case of non-IRS communications up to 8 MHz bandwidth is required to process the maximum size of data. On the other hand, in the case of IRS-assisted communication allocating a minimal bandwidth (up to 2 MHz) all the computation tasks can be processed (task offloading and processing at MEC) within the threshold task completion time. Deploying IRS reduces the bandwidth requirement up to 4 times. Therefore, it is evident that the incorporation of IRS ensures significant reduction of spectrum usage and offers notable spectrum efficiency.\par
Another important point is that compared to the non-IRS communication in which 5 W uplink transmit is considered with 2 W of uplink transmit power the IRS-assisted network performs significantly better. Reducing 40\% uplink transmission power IRS-assisted communication offers remarkable energy efficiency as well.}\par
\vspace{18pt}

\RaggedRight{\textbf{\Large 5.\hspace{10pt} Conclusion}}\\
\vspace{12pt}

\justifying{\noindent The research targeted to analyze the performance of MEC systems under non-IRS and IRS-assisted communication scenarios. In this context, the work reviewed relative works and literature to obtain an insight into prior works. Then, it formulated a system model including the communication and computation to perform the measurement campaign. The measurements are performed through a computer-aided simulation or measurement approach i.e. MATLAB is utilized for the measurements. The research obtained that the deployment of IRS in the MEC ecosystem can significantly improve the overall performance of the system. The deployment of IRS remarkably reduces the spectrum and energy usage hence increasing the spectral and energy efficiency of the network. The authors pretend that the work will enhance the literature on IRS-assisted communication and will be assistive to extend research on relative topics.}\par
\vspace{18pt}

\RaggedRight{\textbf{\Large References}}\\
\begin{enumerate}
\justifying{
    \item	Mohamed I. AlHajri, Nazar T. Ali, and Raed M. Shubair. "Classification of indoor environments for IoT applications: A machine learning approach." IEEE Antennas and Wireless Propagation Letters 17, no. 12 (2018): 2164-2168.
    \item M. I. AlHajri, A. Goian, M. Darweesh, R. AlMemari, R. M. Shubair, L. Weruaga, and A. R. Kulaib. "Hybrid RSS-DOA technique for enhanced WSN localization in a correlated environment." In 2015 International Conference on Information and Communication Technology Research (ICTRC), pp. 238-241. IEEE, 2015.
    \item Mohamed I. AlHajri, Nazar T. Ali, and Raed M. Shubair. "Indoor localization for IoT using adaptive feature selection: A cascaded machine learning approach." IEEE Antennas and Wireless Propagation Letters 18, no. 11 (2019): 2306-2310.
    \item Fahad Belhoul, Raed M. Shubair, and Mohammed E. Al-Mualla. "Modelling and performance analysis of DOA estimation in adaptive signal processing arrays." In ICECS, pp. 340-343. 2003.
    \item Ebrahim M. Al-Ardi, Raed M. Shubair, and Mohammed E. Al-Mualla. "Direction of arrival estimation in a multipath environment: An overview and a new contribution." Applied Computational Electromagnetics Society Journal 21, no. 3 (2006): 226.
    \item R. M. Shubair and A. Al-Merri. "Robust algorithms for direction finding and adaptive beamforming: performance and optimization." In The 2004 47th Midwest Symposium on Circuits and Systems, 2004. MWSCAS'04., vol. 2, pp. II-II. IEEE, 2004.
    \item M. I. AlHajri, N. Alsindi, N. T. Ali, and R. M. Shubair. "Classification of indoor environments based on spatial correlation of RF channel fingerprints." In 2016 IEEE international symposium on antennas and propagation (APSURSI), pp. 1447-1448. IEEE, 2016.
    \item Mohamed AlHajri, Abdulrahman Goian, Muna Darweesh, Rashid AlMemari, Raed Shubair, Luis Weruaga, and Ahmed AlTunaiji. "Accurate and robust localization techniques for wireless sensor networks." arXiv preprint arXiv:1806.05765 (2018).
    \item A. Goian, Mohamed I. AlHajri, Raed M. Shubair, Luis Weruaga, Ahmed Rashed Kulaib, R. AlMemari, and Muna Darweesh. "Fast detection of coherent signals using pre-conditioned root-MUSIC based on Toeplitz matrix reconstruction." In 2015 IEEE 11th International Conference on Wireless and Mobile Computing, Networking and Communications (WiMob), pp. 168-174. IEEE, 2015.
    \item Zhenghua Chen, Mohamed I. AlHajri, Min Wu, Nazar T. Ali, and Raed M. Shubair. "A novel real-time deep learning approach for indoor localization based on RF environment identification." IEEE Sensors Letters 4, no. 6 (2020): 1-4.
    \item Mohamed I. AlHajri, Nazar T. Ali, and Raed M. Shubair. "A machine learning approach for the classification of indoor environments using RF signatures." In 2018 IEEE Global Conference on Signal and Information Processing (GlobalSIP), pp. 1060-1062. IEEE, 2018.
    \item Raed M. Shubair, Abdulrahman S. Goian, Mohamed I. AlHajri, and Ahmed R. Kulaib. "A new technique for UCA-based DOA estimation of coherent signals." In 2016 16th Mediterranean Microwave Symposium (MMS), pp. 1-3. IEEE, 2016.
    \item M. I. AlHajri, R. M. Shubair, L. Weruaga, A. R. Kulaib, A. Goian, M. Darweesh, and R. AlMemari. "Hybrid method for enhanced detection of coherent signals using circular antenna arrays." In 2015 IEEE International Symposium on Antennas and Propagation \& USNC/URSI National Radio Science Meeting, pp. 1810-1811. IEEE, 2015.
    \item WafaNjima, Marwa Chafii, ArseniaChorti, Raed M. Shubair, and H. Vincent Poor. "Indoor localization using data augmentation via selective generative adversarial networks." IEEE Access 9 (2021): 98337-98347.
    \item WafaNjima, Marwa Chafii, and Raed M. Shubair. "Gan based data augmentation for indoor localization using labeled and unlabeled data." In 2021 International Balkan Conference on Communications and Networking (BalkanCom), pp. 36-39. IEEE, 2021.
    \item Mohamed I. AlHajri, Nazar T. Ali, and Raed M. Shubair. "A cascaded machine learning approach for indoor classification and localization using adaptive feature selection." AI for Emerging Verticals: Human-robot computing, sensing and networking (2020): 205.
    \item Mohamed I. AlHajri, Raed M. Shubair, and Marwa Chafii. "Indoor Localization Under Limited Measurements: A Cross-Environment Joint Semi-Supervised and Transfer Learning Approach." In 2021 IEEE 22nd International Workshop on Signal Processing Advances in Wireless Communications (SPAWC), pp. 266-270. IEEE, 2021.
    \item Raed M. Shubair and Hadeel Elayan. "In vivo wireless body communications: State-of-the-art and future directions." In 2015 Loughborough Antennas \& Propagation Conference (LAPC), pp. 1-5. IEEE, 2015.
    \item Hadeel Elayan, Raed M. Shubair, and Asimina Kiourti. "Wireless sensors for medical applications: Current status and future challenges." In 2017 11th European Conference on Antennas and Propagation (EUCAP), pp. 2478-2482. IEEE, 2017.
    \item Hadeel Elayan, Raed M. Shubair, and Asimina Kiourti. "Wireless sensors for medical applications: Current status and future challenges." In 2017 11th European Conference on Antennas and Propagation (EUCAP), pp. 2478-2482. IEEE, 2017.
    \item Hadeel Elayan, Raed M. Shubair, Josep Miquel Jornet, and Raj Mittra. "Multi-layer intrabody terahertz wave propagation model for nanobiosensing applications." Nano communication networks 14 (2017): 9-15.
    \item Hadeel Elayan, Pedram Johari, Raed M. Shubair, and Josep Miquel Jornet. "Photothermal modeling and analysis of intrabody terahertz nanoscale communication." IEEE transactions on nanobioscience 16, no. 8 (2017): 755-763.
    \item Rui Zhang, Ke Yang, Akram Alomainy, Qammer H. Abbasi, Khalid Qaraqe, and Raed M. Shubair. "Modelling of the terahertz communication channel for in-vivo nano-networks in the presence of noise." In 2016 16th Mediterranean Microwave Symposium (MMS), pp. 1-4. IEEE, 2016.
    \item Samar Elmeadawy and Raed M. Shubair. "6G wireless communications: Future technologies and research challenges." In 2019 international conference on electrical and computing technologies and applications (ICECTA), pp. 1-5. IEEE, 2019.
    \item Hadeel Elayan, Raed M. Shubair, and Asimina Kiourti. "On graphene-based THz plasmonic nano-antennas." In 2016 16th mediterranean microwave symposium (MMS), pp. 1-3. IEEE, 2016.
    \item Hadeel Elayan, Cesare Stefanini, Raed M. Shubair, and Josep Miquel Jornet. "End-to-end noise model for intra-body terahertz nanoscale communication." IEEE transactions on nanobioscience 17, no. 4 (2018): 464-473.
    \item Hadeel Elayan, Raed M. Shubair, Akram Alomainy, and Ke Yang. "In-vivo terahertz em channel characterization for nano-communications in wbans." In 2016 IEEE International Symposium on Antennas and Propagation (APSURSI), pp. 979-980. IEEE, 2016.
    \item Hadeel Elayan, Raed M. Shubair, and Josep M. Jornet. "Bio-electromagnetic thz propagation modeling for in-vivo wireless nanosensor networks." In 2017 11th European Conference on Antennas and Propagation (EuCAP), pp. 426-430. IEEE, 2017.
    \item Hadeel Elayan, Raed M. Shubair, and Josep M. Jornet. "Characterising THz propagation and intrabody thermal absorption in iWNSNs." IET Microwaves, Antennas \& Propagation 12, no. 4 (2018): 525-532.
    \item S. Elmeadawy, and R. M. Shubair. "Enabling technologies for 6G future wireless communications: Opportunities and challenges. arXiv 2020." arXiv preprint arXiv:2002.06068.
    \item Hadeel Elayan, Cesare Stefanini, Raed M. Shubair, and Josep M. Jornet. "Stochastic noise model for intra-body terahertz nanoscale communication." In Proceedings of the 5th ACM International Conference on Nanoscale Computing and Communication, pp. 1-6. 2018.
    \item Abdul Karim Gizzini, Marwa Chafii, Ahmad Nimr, Raed M. Shubair, and Gerhard Fettweis. "Cnn aided weighted interpolation for channel estimation in vehicular communications." IEEE Transactions on Vehicular Technology 70, no. 12 (2021): 12796-12811.
    \item Nishtha Chopra, Mike Phipott, Akram Alomainy, Qammer H. Abbasi, Khalid Qaraqe, and Raed M. Shubair. "THz time domain characterization of human skin tissue for nano-electromagnetic communication." In 2016 16th Mediterranean Microwave Symposium (MMS), pp. 1-3. IEEE, 2016.
    \item Hadeel Elayan, Raed M. Shubair, Josep M. Jornet, Asimina Kiourti, and Raj Mittra. "Graphene-Based Spiral Nanoantenna for Intrabody Communication at Terahertz." In 2018 IEEE International Symposium on Antennas and Propagation \& USNC/URSI National Radio Science Meeting, pp. 799-800. IEEE, 2018.
    \item Taki Hasan Rafi, Raed M. Shubair, Faisal Farhan, Md Ziaul Hoque, and Farhan Mohd Quayyum. "Recent Advances in Computer-Aided Medical Diagnosis Using Machine Learning Algorithms with Optimization Techniques." IEEE Access (2021).
    \item Abdul Karim Gizzini, Marwa Chafii, Shahab Ehsanfar, and Raed M. Shubair. "Temporal Averaging LSTM-based Channel Estimation Scheme for IEEE 802.11 p Standard." arXiv preprint arXiv:2106.04829 (2021).
    \item Menna El Shorbagy, Raed M. Shubair, Mohamed I. AlHajri, and Nazih Khaddaj Mallat. "On the design of millimetre-wave antennas for 5G." In 2016 16th Mediterranean Microwave Symposium (MMS), pp. 1-4. IEEE, 2016.
    \item Malak Y. ElSalamouny, and Raed M. Shubair. "Novel design of compact low-profile multi-band microstrip antennas for medical applications." In 2015 loughborough antennas \& propagation conference (LAPC), pp. 1-4. IEEE, 2015.
    \item Raed M. Shubair, Amna M. AlShamsi, Kinda Khalaf, and Asimina Kiourti. "Novel miniature wearable microstrip antennas for ISM-band biomedical telemetry." In 2015 Loughborough Antennas \& Propagation Conference (LAPC), pp. 1-4. IEEE, 2015.
    \item Muhammad S. Khan, Syed A. Naqvi, Adnan Iftikhar, Sajid M. Asif, Adnan Fida, and Raed M. Shubair. "A WLAN band‐notched compact four element UWB MIMO antenna." International Journal of RF and Microwave Computer‐Aided Engineering 30, no. 9 (2020): e22282.
    \item Saad Alharbi, Raed M. Shubair, and Asimina Kiourti. "Flexible antennas for wearable applications: Recent advances and design challenges." (2018): 484-3.
    \item M. S. Khan, F. Rigobello, Bilal Ijaz, E. Autizi, A. D. Capobianco, R. Shubair, and S. A. Khan. "Compact 3‐D eight elements UWB‐MIMO array." Microwave and Optical Technology Letters 60, no. 8 (2018): 1967-1971.
    \item R. Karli, H. Ammor, R. M. Shubair, M. I. AlHajri, and A. Hakam. "Miniature Planar Ultra-Wide-Band Microstrip Patch Antenna for Breast Cancer Detection." Skin 1 (2016): 39.
    \item Ala Eldin Omer, George Shaker, Safieddin Safavi-Naeini, Kieu Ngo, Raed M. Shubair, Georges Alquié, Frédérique Deshours, and Hamid Kokabi. "Multiple-cell microfluidic dielectric resonator for liquid sensing applications." IEEE Sensors Journal 21, no. 5 (2020): 6094-6104.
    \item Muhammad Saeed Khan, Adnan Iftikhar, Raed M. Shubair, Antonio-Daniele Capobianco, Sajid Mehmood Asif, Benjamin D. Braaten, and Dimitris E. Anagnostou. "Ultra-compact reconfigurable band reject UWB MIMO antenna with four radiators." Electronics 9, no. 4 (2020): 584.
    \item Muhammad S. Khan, Adnan Iftikhar, Raed M. Shubair, Antonio D. Capobianco, Benjamin D. Braaten, and Dimitris E. Anagnostou. "A four element, planar, compact UWB MIMO antenna with WLAN band rejection capabilities." Microwave and Optical Technology Letters 62, no. 10 (2020): 3124-3131.
    \item Yazan Al-Alem, Ahmed A. Kishk, and Raed M. Shubair. "Enhanced wireless interchip communication performance using symmetrical layers and soft/hard surface concepts." IEEE Transactions on Microwave Theory and Techniques 68, no. 1 (2019): 39-50.
    \item Yazan Al-Alem, Ahmed A. Kishk, and Raed M. Shubair. "One-to-two wireless interchip communication link." IEEE Antennas and Wireless Propagation Letters 18, no. 11 (2019): 2375-2378.
    \item Yazan Al-Alem, Raed M. Shubair, and Ahmed Kishk. "Efficient on-chip antenna design based on symmetrical layers for multipath interference cancellation." In 2016 16th Mediterranean Microwave Symposium (MMS), pp. 1-3. IEEE, 2016.
    \item Asimina Kiourti, and Raed M. Shubair. "Implantable and ingestible sensors for wireless physiological monitoring: a review." In 2017 IEEE International Symposium on Antennas and Propagation \& USNC/URSI National Radio Science Meeting, pp. 1677-1678. IEEE, 2017.
    \item Yazan Al-Alem, Raed M. Shubair, and Ahmed Kishk. "Clock jitter correction circuit for high speed clock signals using delay units a nd time selection window." In 2016 16th Mediterranean Microwave Symposium (MMS), pp. 1-3. IEEE, 2016.
    \item Melissa Eugenia Diago-Mosquera, Alejandro Aragón-Zavala, Fidel Alejandro Rodríguez-Corbo, Mikel Celaya-Echarri, Raed M. Shubair, and Leyre Azpilicueta. "Tuning Selection Impact on Kriging-Aided In-Building Path Loss Modeling." IEEE Antennas and Wireless Propagation Letters 21, no. 1 (2021): 84-88.
    \item Mikel Celaya-Echarri, Leyre Azpilicueta, Fidel Alejandro Rodríguez-Corbo, Peio Lopez-Iturri, Victoria Ramos, Mohammad Alibakhshikenari, Raed M. Shubair, and Francisco Falcone. "Towards Environmental RF-EMF Assessment of mmWave High-Node Density Complex Heterogeneous Environments." Sensors 21, no. 24 (2021): 8419.
    \item Yazan Al-Alem, Ahmed A. Kishk, and Raed Shubair. "Wireless chip to chip communication link budget enhancement using hard/soft surfaces." In 2018 IEEE Global Conference on Signal and Information Processing (GlobalSIP), pp. 1013-1014. IEEE, 2018.
    \item Yazan Al-Alem, Yazan, Ahmed A. Kishk, and Raed M. Shubair. "Employing EBG in Wireless Inter-chip Communication Links: Design and Performance." In 2020 IEEE International Symposium on Antennas and Propagation and North American Radio Science Meeting, pp. 1303-1304. IEEE, 2020.
    \item S. Elmeadawy and R. M. Shubair, "6G Wireless Communications: Future Technologies and Research Challenges," 2019 International Conference on Electrical and Computing Technologies and Applications (ICECTA), 2019.
    \item E. A. Kadir, R. Shubair, S. K. Abdul Rahim, M. Himdi, M. R. Kamarudin and S. L. Rosa, "B5G and 6G: Next Generation Wireless Communications Technologies, Demand and Challenges," 2021 International Congress of Advanced Technology and Engineering (ICOTEN), 2021, pp. 1-6.
    \item W. Jiang, B. Han, M. A. Habibi and H. D. Schotten, "The Road Towards 6G: A Comprehensive Survey," in \textit{IEEE Open Journal of the Communications Society}, vol. 2, pp. 334-366, 2021.
    \item Y. Mao, C. You, J. Zhang, K. Huang and K. B. Letaief, "A Survey on Mobile Edge Computing: The Communication Perspective," in IEEE Communications Surveys \& Tutorials, vol. 19, no. 4, pp. 2322-2358, Fourthquarter 2017.
    \item F. Spinelli and V. Mancuso, "Toward Enabled Industrial Verticals in 5G: A Survey on MEC-Based Approaches to Provisioning and Flexibility," in IEEE Communications Surveys \& Tutorials, vol. 23, no. 1, pp. 596-630, Firstquarter 2021.
    \item C. Pan et al., "Reconfigurable Intelligent Surfaces for 6G Systems: Principles, Applications, and Research Directions," in IEEE Communications Magazine, vol. 59, no. 6, pp. 14-20, June 2021.
    \item Q. Wu, S. Zhang, B. Zheng, C. You and R. Zhang, "Intelligent Reflecting Surface-Aided Wireless Communications: A Tutorial," in IEEE Transactions on Communications, vol. 69, no. 5, pp. 3313-3351, May 2021.
    \item X. Pei et al., "RIS-Aided Wireless Communications: Prototyping, Adaptive Beamforming, and Indoor/Outdoor Field Trials," in IEEE Transactions on Communications, vol. 69, no. 12, pp. 8627-8640, Dec. 2021.
    \item F. Zhou, C. You and R. Zhang, "Delay-Optimal Scheduling for IRS-Aided Mobile Edge Computing," in IEEE Wireless Communications Letters, vol. 10, no. 4, pp. 740-744, April 2021.
    \item Q. Wang, F. Zhou, H. Hu and R. Q. Hu, "Energy-Efficient Design for IRS-Assisted MEC Networks with NOMA," 2021 13th International Conference on Wireless Communications and Signal Processing (WCSP), 2021, pp. 1-6.
    \item F. Zhou, C. You and R. Zhang, "Delay-Optimal Scheduling for IRS-Aided Mobile Edge Computing," in IEEE Wireless Communications Letters, vol. 10, no. 4, pp. 740-744, April 2021.
    \item Z. Chu, P. Xiao, M. Shojafar, D. Mi, J. Mao and W. Hao, "Intelligent Reflecting Surface Assisted Mobile Edge Computing for Internet of Things," in IEEE Wireless Communications Letters, vol. 10, no. 3, pp. 619-623, March 2021.
    \item H. Zhang, X. He, Q. Wu and H. Dai, "Spectral Graph Theory Based Resource Allocation for IRS-Assisted Multi-Hop Edge Computing," IEEE INFOCOM 2021 - IEEE Conference on Computer Communications Workshops (INFOCOM WKSHPS), 2021, pp. 1-6.
    \item T. Bai, C. Pan, Y. Deng, M. Elkashlan, A. Nallanathan and L. Hanzo, "Latency Minimization for Intelligent Reflecting Surface Aided Mobile Edge Computing," in IEEE Journal on Selected Areas in Communications, vol. 38, no. 11, pp. 2666-2682, Nov. 2020.
    \item S. Mao et al., "Computation Rate Maximization for Intelligent Reflecting Surface Enhanced Wireless Powered Mobile Edge Computing Networks," in IEEE Transactions on Vehicular Technology, vol. 70, no. 10, pp. 10820-10831, Oct. 2021.
    \item C. Sun, W. Ni, Z. Bu and X. Wang, "Energy Minimization for Intelligent Reflecting Surface-Assisted Mobile Edge Computing," in IEEE Transactions on Wireless Communications, Feb. 2022.
    \item T. Mir, L. Dai, Y. Yang, W. Shen and B. Wang, "Optimal FemtoCell Density for Maximizing Throughput in 5G Heterogeneous Networks under Outage Constraints," 2017 IEEE 86th Vehicular Technology Conference (VTC-Fall), Toronto, ON, Canada, 2017, pp. 1-5.
    \item W. Tang et al., "Wireless Communications With Reconfigurable Intelligent Surface: Path Loss Modeling and Experimental Measurement," in IEEE Transactions on Wireless Communications, vol. 20, no. 1, pp. 421-439, Jan. 2021.
    \item Y. Dai, D. Xu, S. Maharjan and Y. Zhang, "Joint Computation Offloading and User Association in Multi-Task Mobile Edge Computing," in IEEE Transactions on Vehicular Technology, vol. 67, no. 12, pp. 12313-12325, Dec. 2018.
    }
\end{enumerate}

\end{document}